\documentclass[useAMS]{mn2e} 
 
\usepackage{graphicx} 
\usepackage{txfonts} 
\newcommand\aj{AJ} 
\newcommand\apj{ApJ}

\newcommand\aap{A\&A} 
\newcommand\mnras{MNRAS} 
\newcommand\apjl{ApJ} 
\newcommand\pasp{PASP}

\title[]{Gaia unveils the kinematics of multiple stellar populations in 47\,Tucanae}
\author[] 
       {A.\,P.\,Milone$^{1}$,
         A.\,F.\,Marino$^{2}$,
         A.\,Mastrobuono-Battisti$^{3}$,
         E.\,P.\,Lagioia$^{1}$
        \\ 
$^{1}$Dipartimento di Fisica e Astronomia ``Galileo Galilei'', Univ. di Padova, Vicolo dell'Osservatorio 3, Padova, IT-35122\\
$^{2}$Research School of Astronomy \& Astrophysics, Australian National University, Canberra, ACT 2611, Australia \\
$^{3}$Max-Planck Instiut f\"{u}r Astronomie, K\"{o}nigstuhl 17, D-69117 Heidelberg, Germany\\}
\begin{document} 
\date{Accepted 2018 July 9. Received 2018 July 6; in original form 2018 May 16}
 
\pagerange{\pageref{firstpage}--\pageref{lastpage}} \pubyear{2017} 
 
\maketitle 
\label{firstpage}

\begin{abstract}
  The series of events, which occurred at high redshift and originated multiple stellar populations in Globular Clusters (GCs) are still poorly understood. Theoretical work suggests that the present-day dynamics of stars in nearby GCs, including the rotation and velocity dispersion, may retain important clues on the formation of multiple populations.

So far, the dynamics of multiple populations have been investigated either from radial velocities of a relatively-small sample of stars, or from relative proper motions  of stars in the small field of view provided by the {\it Hubble Space Telescope}. In this context, Gaia provides the unique opportunity to investigate the dynamics of thousands GC stars over a wide field of view.  

  For the first time, we combine Gaia DR2 proper motions and multi-band photometry to study the internal motions of the two main stellar populations of 47\,Tucanae in a wide field of view.
  We confirm that this cluster exhibits high rotation on the plane of the sky and find that both stellar generations share similar rotation patters.
  Second-generation stars show stronger anisotropies and smaller tangential-velocity dispersion than the first generation, while there is no significant difference between their radial-velocity dispersion profiles. 
  We discuss the impact of these results in the context of the formation scenarios for multiple stellar populations in GCs.
\end{abstract} 
 
 
\section{Introduction}\label{sec:intro}
{\it N}-body simulations have shown that the internal dynamics of the distinct stellar populations in Globular Clusters (GCs) provide strong constraints on the formation scenarios of their multiple populations. Specifically, the present-day rotation and the velocity-dispersion profile of first-generation (1G) and second-generation (2G) stars would be related to the formation of 2G stars and their initial configuration (Mastrobuono-Battisti \& Perets 2013, 2016; Vesperini et al.\,2013; H{\'e}nault-Brunet et al.\,2015).

The rotation of multiple populations in GCs has been poorly investigated to date.   Nearly all the previous studies were indeed based on radial velocities and were limited by small sample sizes.
 $\omega$\,Centauri is a remarkable exception. Indeed, from the analysis of {\it Hubble Space Telescope} ({\it HST\,}) proper motions, Bellini et al.\,(2018) show that 1G stars have excess systemic rotation in the plane of the sky with respect to 2G stars.

The largest stellar sample based on radial velocities was used by Pancino et al.\,(2007) who analyzed 650 stars in $\omega$ Centauri and studied the rotation along the line of sight of the three main sub-populations of metal-poor, metal-intermediate, and metal-rich stars.
They concluded that the three populations are all compatible with having the same rotational pattern, in contrast with previous finding by Norris et al.\,(1997) who showed that, while the majority of stars in $\omega$ Centauri exhibits strong rotation, the most metal-rich stars do not show any sign of rotation.

 Bellazzini et al.\,(2012) analyzed the radial velocities of 1,981 stars in 20 GCs and  did not find any relation between the presence of multiple populations and the rotation within each cluster.

In contrast, evidence that stars with extreme abundance of light elements exhibit different rotational patters than the remaining cluster stars comes from the study of 113 red-giant branch (RGB) stars in M\,13 (Cordero et al.\,2017).
These authors concluded that the 24 analyzed Na-enhanced and extremely O-depleted stars exhibit faster rotation than the other stars.

The velocity dispersion of multiple stellar populations was typically studied by using spectroscopy.  
Bellazzini et al.\,(2012), concluded that in 17 out of 20 analyzed clusters, the stars with different light-elements abundance have similar velocity-dispersion profiles.  The massive GCs NGC\,6388, NGC\,6441, and NGC\,2808, where sodium-rich stars seem to have a slightly lower line-of-sight velocity dispersion, are possible exceptions to this rule.
 Similarly, Marino et al.\,(2014) show that the two main populations of stars with different slow neutron-capture element abundance of NGC\,1851 have similar radial-velocity dispersion. 

In a few cases, high-precision proper motions from  {\it HST\,} allowed to extend the investigation of the velocity dispersion to a large number of thousands stars.
  Four GCs have been analyzed to date with {\it HST}, namely NGC\,104 (47\,Tuc) NGC\,2808, and $\omega$\,Centauri (Richer et al.\,2013; Bellini et al.\,2015, 2018)  and  NGC\,362 (Libralato et al.\,2018). In all the cases the stars with extreme helium abundances have more radially-anisotropic velocity distribution in the plane of the sky. For NGC\,362 this result is significant at 2.2$\sigma$ level.
 Unfortunately, these studies are limited to the relatively-small field of view covered by the {\it HST} cameras.

In this context, 47\,Tuc is an interesting case, which reveals an high degree of dynamical complexity. Specifically, {\it HST\,} proper motions have revealed that this cluster exhibits high internal rotation (Anderson \& King 2003; Bellini et al.\,2017; Gaia collaboration et al.\,2018a; Bianchini et al.\,2018) and significant radial anisotropy in the external region (Richer et al.\,2013; Bellini et al.\,2017).

Since the seventies, 47\,Tuc has been widely studied in the context of stellar populations both spectroscopically and photometrically (e.g.\,Norris \& Freeman 1979, 1982; Anderson et al.\,2009; Cordero et al.\,2014; Marino et al.\,2016; Wang et al.\,2017).
 High-precision {\it HST} and ground-based photometry has revealed that its color-magnitude diagram (CMD) is formed of two main discrete sequences of stars that can be followed along the various evolutionary stages from the main sequence (MS) to the asymptotic giant branch (Milone et al.\,2012; Piotto et al.\,2015).
 These sequences correspond to a first stellar generation (1G), which is formed of stars with a chemical composition similar to that of halo field stars at similar metallicity, and to the a second generation (2G) of stars enhanced in helium, nitrogen and sodium and depleted in carbon and oxygen. Both groups of 1G and 2G stars host sub populations (e.g.\,Marino et al.\,2016; Milone et al.\,2017). 

 The two main populations of 47\,Tuc exhibit different radial distributions, with 2G stars being significantly more-centrally concentrated than 1G stars (Norris \& Freeman 1979; Milone et al.\,2012; Cordero et al.\,2014).
Clearly, the stellar populations of 47\,Tuc are not well mixed and could still retain information on their star-formation history.
 
 The first investigation of the dynamics of stellar populations of 47\,Tuc was provided by Richer et al.\,(2013) who used {\it HST} photometry and proper motions.
 These authors have divided MS stars into four groups, which presumably correspond to stellar populations with different chemical abundances, and found that the anisotropy in the proper-motion distribution correlates with stellar colors.
 Specifically, the bluest stars exhibit the most-pronounced proper-motion anisotropy while red stars show isotropic proper motions. 
 This finding corroborates similar conclusion by Ku{\v c}inskas et al.\,(2014), who analyzed the spectra of 101 stars of 47\,Tuc and detected a significant correlation between the velocity dispersion along the line of sight and the O and Na abundance.

In this work we combine wide-field ground-based photometry and stellar proper motions from Gaia data release 2 (DR2, Gaia collaboration et al.\,2018b) to further investigate the internal dynamics of multiple stellar populations in 47\,Tuc. For the first time, this analysis will be extended to a large field of view.

The paper is organized as follows. In Section~\ref{sec:data} we describe the data and present the photometric diagrams of 47\,Tuc. The dynamics of 1G and 2G stars are investigated in Section~\ref{sec:dynamics}. Finally, a summary of the results and a discussion is provided in Section~\ref{sec:summary}. 

\section{Data} \label{sec:data} 
 We combined ground-based wide-field photometry and proper motions from Gaia DR2 to investigate the internal dynamics of stellar populations in 47\,Tuc.
 To identify multiple populations in the CMD we used $U$, $B$, $V$, $I$ photometry derived from 856 images collected with various facilities, including the Wide-Field Imager of the ESO/MPI telescope and the 1.5 m telescope at Cerro Tololo Inter-American Observatory (Stetson 2000). These images have been reduced by Peter Stetson by using the methods and the computer programs by Stetson (2005) and are calibrated on the photometric system by Landolt (2002). Details on this dataset and on the data reduction are provided by Bergbush \& Stetson (2009).

 This photometric catalog has been used in previous studies on multiple populations showing that two distinct groups of 1G and 2G stars are clearly visible along the RGB and the horizontal branch (HB) of 47\,Tuc (Milone et al.\,2012; Monelli et al.\,2013; Marino et al. 2016). The presence of two populations is evident in various diagrams involving the $U$, $B$, $I$ filters, like the $U$-$B$ vs.\,$B-I$ two-color diagram and the $B$ vs.\,($U-B+I$) or $B$ vs.\,$U-2 \cdot B+I$=$C_{\rm U,B,I}$ pseudo CMDs.
 Stetson's catalog was matched with the Gaia DR2 one and only stars for which both $U$, $B$, $V$, $I$ photometry and Gaia proper motions are available are used in this paper.

 Moreover, we excluded stars with poor Gaia astrometry.  To do this, we first used the parameter released with Gaia DR2 `astrometric\_gof\_al', which is indicative of the goodness of fit statistic of the astrometric solution for the source in the along-scan direction (see Gaia Collaboration et al.\,2018a for details). When we plot this parameter as a function of the g-band magnitude of Gaia, most of the stars exhibit a clear trend. To exclude the ouliers from the analysis we followed a procedure similar to the one described by Milone et al.\,(2009, see their Sect.~2.1). Briefly, we divided the magnitude interval covered by the RGB stars of 47\,Tuc into bins of 0.5 mag. For each bin, we calculated the median magnitude, the median value of the astrometric\_gof\_al  parameter and the corresponding 68.27$^{o}$ percentile ($\sigma$). We excluded from the analysis all the stars that exceed the median values by $N=5$ times $\sigma$ and iterated this procedure until two subsequent values of the median and the $\sigma$ values differ by less than 0.01. 
   In addition, we excluded stars with proper-motion uncertainties larger than 0.15 mas yr$^{-1}$. 
   Indeed, we noticed that the bulk of our sample of relatively-bright RGB and HB stars have uncertainties below this value\footnote{  We carefully checked that our results are not affected by the adopted selection criteria. To do this, we verified that by using values of $2 \sigma$ and $10 \sigma$ and by excluding stars with  with proper motion uncertainties larger than 0.08 mas yr$^{-1}$ and 0.30 mas yr$^{-1}$ the conclusions of the paper remain unchanged. Similarly,we repeated the analysis by excluding stars with correlation coefficients that differ by more than $\pm 1 \sigma$ from the median value.
       We find that all the conclusions remain unchanged, thus demonstrating that there is no evidence for any significant bias connected to the correlation coefficients between the proper motions along the right ascension and the declination.  }.
 The final sample comprises 3,276 cluster members between 0.8 and 18.0 arcmin from the cluster center, including 1,208 1G stars and 2,068 2G stars.

\begin{centering} 
\begin{figure*} 
  \includegraphics[width=11.0cm]{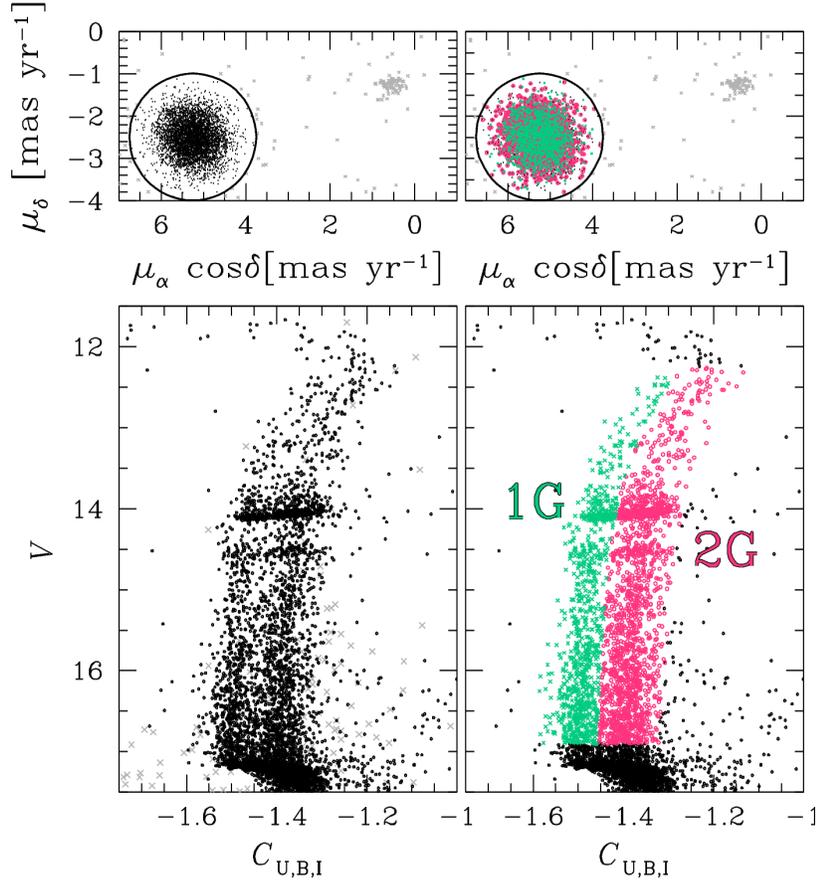}
  \caption{$V$ vs.\,$C_{\rm U,B,I}$ pseudo-CMDs (lower panels) and vector-point diagrams of stars in 47\,Tuc (upper panels). The black circle in the vector point diagram separates cluster members from field stars, which are colored black and gray, respectively in the left panels. Note the Small-Magellanic Cloud stars clustering around ($\mu_{\alpha}cos{\delta};\mu_{\delta}$)$\sim$($-0.5;-1.2$).  In the right-panels we used aqua and magenta colors to represent 1G and 2G stars.}
 \label{fig:cmd} 
\end{figure*} 
\end{centering}

 The left panels of Figure~\ref{fig:cmd} show the $V$ vs.\,$C_{\rm U,B,I}$ pseudo-CMD of 47\,Tuc zoomed around the RGB and the HB (bottom) and the vector-point diagram of proper motion (top). We used the black circle to separate bona-fide cluster members from field stars, which are represented with black dots and grey crosses, respectively.  We verified that our criterion is consistent with the membership selection by Gaia Collaboration et al.\,(2018c) for the sample of stars analyzed in this paper.
 The two main RGBs and HBs of 47\,Tuc, which we have widely investigated in previous papers (e.g.\,Milone et al.\,2012; Monelli et al.\,2013; Marino et al.\,2016), are clearly visible in this diagram and we used aqua and magenta colors to mark 1G and 2G stars, respectively, in the right-panel diagrams. 
 
 In our analysis we exploit the position of the cluster center, the values of core and half-light radius and the cluster distance provided by the Harris (1996, updated as in 2010) catalog. Their values are listed in Table~1.

\begin{table}
 \begin{centering}
  \caption{Parameters of 47\,Tuc used in this paper. The average proper motions are derived by using Gaia DR2 data. The remaining parameters are taken from the catalog by Harris\,(1996, updates as in 2010). }
 \begin{tabular}{lc}
   \hline
   \hline
   R.\,A.\,(J2000) &   00 24 05.67  \\  
   DEC.\,(J2000)   &$-$72 04 52.6   \\  
   Core radius     &   0.36 arcmin  \\  
   Half-light radius &  3.17 arcmin \\  
   Distance &  4.5 kpc              \\  
   $\mu_{\alpha} cos{\delta}$ & 5.25$\pm$0.01 mas yr$^{-1}$ \\ 
   $\mu_{\delta}$      & $-$2.49$\pm$0.01 mas yr$^{-1}$ \\ 
   \hline
   \hline
 \end{tabular}\\
 \end{centering}
 \label{tab:parametri}
 \end{table}


\section{Dynamics of multiple stellar populations}
\label{sec:dynamics}

The high-precision proper motions from Gaia DR2 allow the investigation of stellar dynamics within 47\,Tuc.
In this section, we use the proper motions of each group of 1G and 2G stars selected in Figure~\ref{fig:cmd} to analyze the rotation in the plane of the sky and the velocity dispersion of each sub-population.

\subsection{Rotation on the plane of the sky}
\begin{centering} 
\begin{figure*} 
  \includegraphics[height=7.5cm]{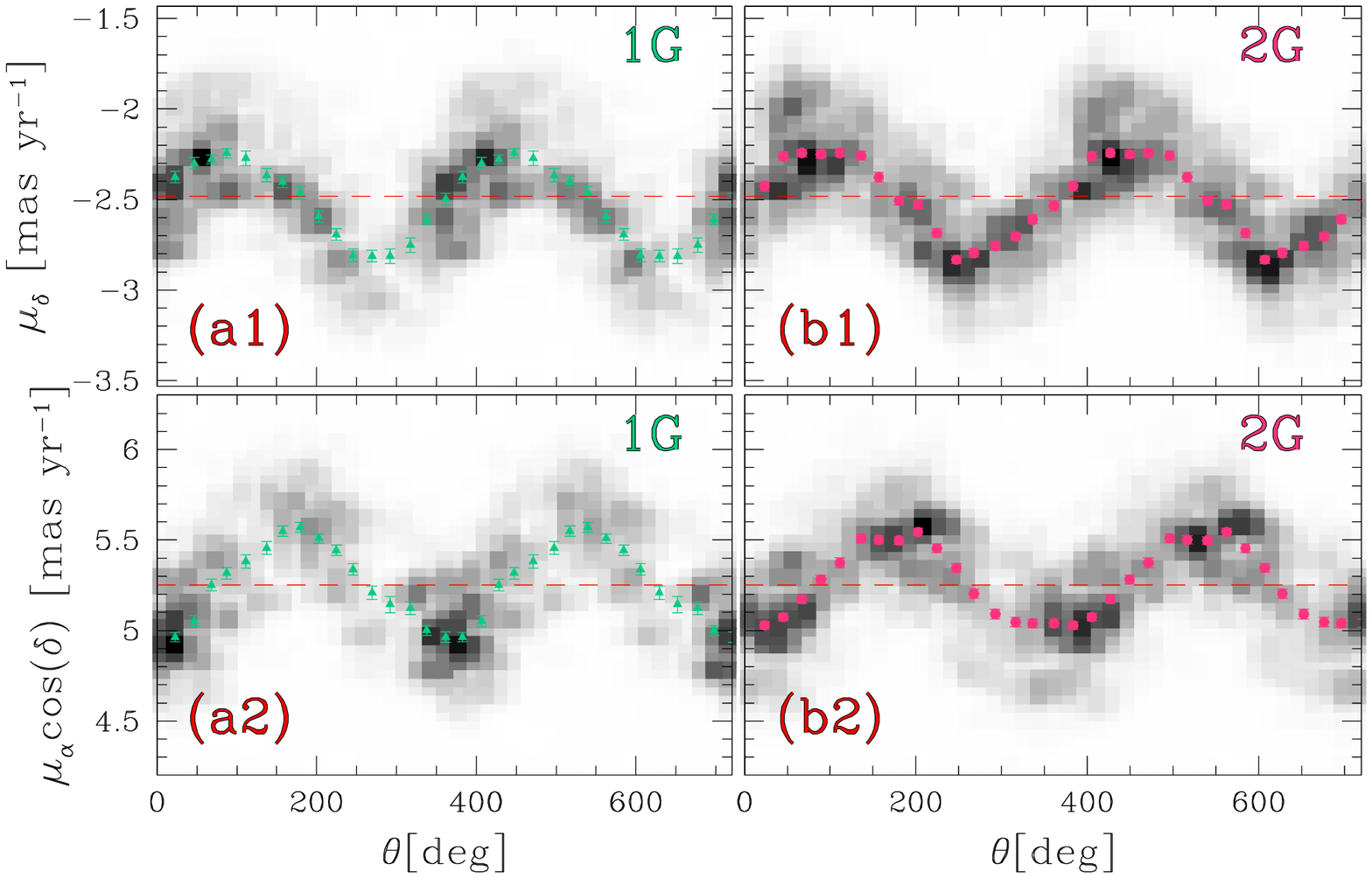}
  \includegraphics[height=7.5cm]{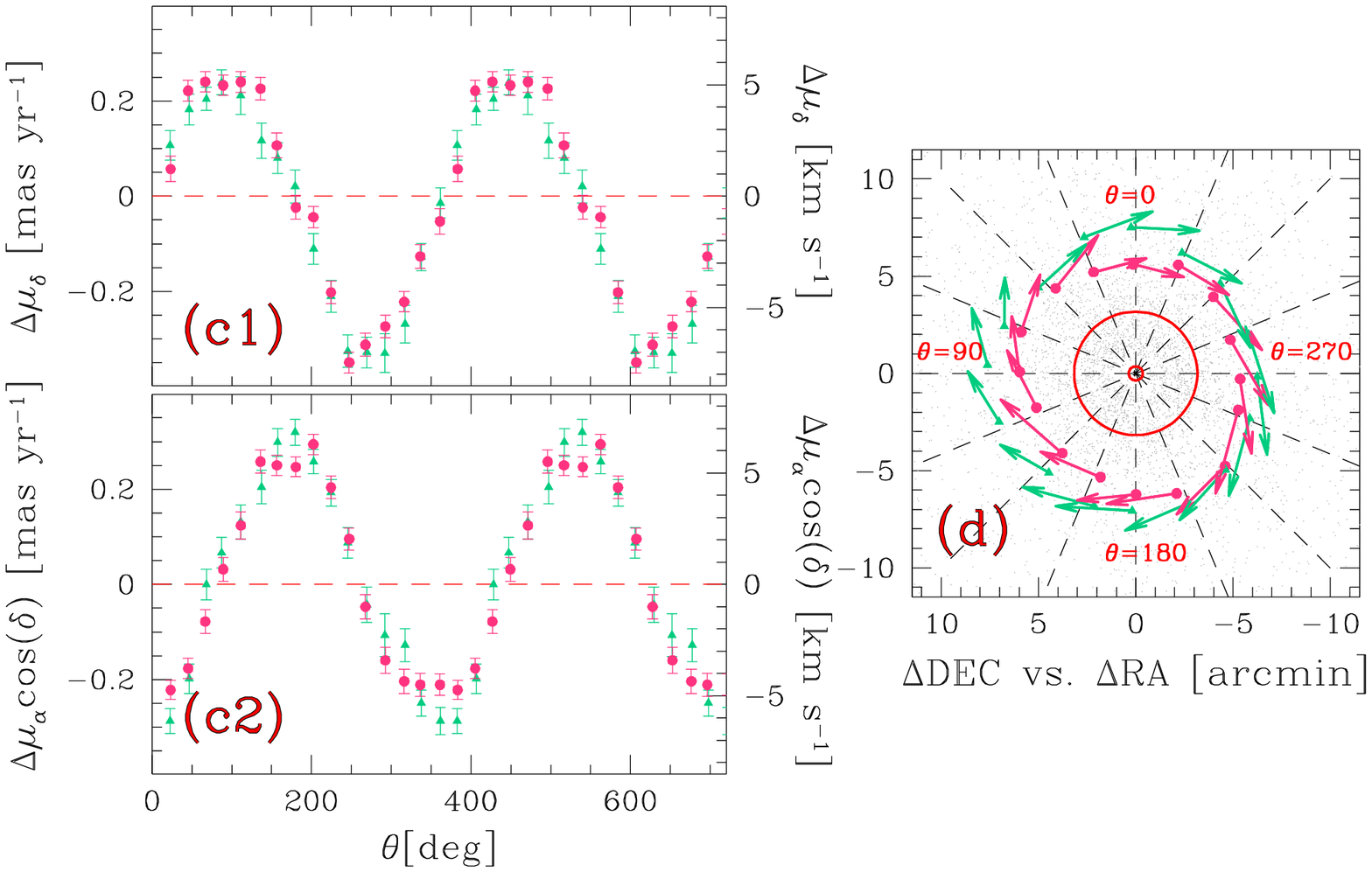}
  \caption{Panels a1-a2, b1-b2 show $\mu_{\alpha} cos{\delta}$ and $\mu_{\delta}$ as a function of the position angle, $\theta$, for 1G and 2G stars, respectively. The gray levels in the panels indicate the density of stars in each population, while the red horizontal dashed lines mark the average proper motions of 47\,Tuc.
The average rotation of 1G and 2G stars, with respect to the average cluster motion is plotted in the panels c1 and c2. 
    Aqua and magenta colors indicate the average motion of 1G and 2G stars, respectively, in different intervals of $\theta$.  
Panel d shows the rotation map of 47\,Tuc. The gray points mark the relative position of the analyzed stars with respect to the cluster center. The core radius and the half-light radius are marked with red circles. The arrows indicate the mean motion in 250,000 years for 1G and 2G stars in the 16 slices delimited by dashed lines.}
 \label{fig:rotazione} 
\end{figure*} 
\end{centering}

To investigate the rotation of the two main populations of 47\,Tuc, we first show in the upper panels of Figure~\ref{fig:rotazione} the density of 1G and 2G stars in the $\mu_{\alpha} cos{\delta}$ vs.\,$\theta$ and $\mu_{\delta}$ vs.\,$\theta$ planes, where $\theta$ is the position angle. The sinusoidal patterns of 1G and 2G stars clearly indicate that both populations exhibit significant rotation on the plane of sky.

We defined a grid of 16 values of $\theta$, ranging from 0$^{\rm o}$ to 360$^{\rm o}$ in steps of 22.5$^{\rm o}$, and associated to each value of $\theta$ a circular sector with an arc length of 45$^{o}$. Then we calculated the median values of $\mu_{\alpha} cos{\delta}$ and $\mu_{\delta}$ of 1G and 2G stars in each circular sector.
 The median points are associated with the mean coordinates of the analyzed stars and are represented with aqua and magenta points for 1G and 2G stars, respectively, in the upper panels Figure~\ref{fig:rotazione}.   

 In the panels c1 and c2 of Figure~\ref{fig:rotazione} we compare the median motions of 1G and 2G stars in the 16 circular sectors after subtracting the median motions of all the analyzed stars\footnote{We find that the median proper motions of all the analyzed stars are $\mu_{\alpha} cos{\delta}$=5.25$\pm$0.01 mas yr$^{-1}$ and $\mu_{\delta}$=$-$2.49$\pm$0.01 mas yr$^{-1}$, and is consistent within 2-$\sigma$ with the determination by Gaia Collaboration et al.\,(2018c, see their Table~C.1).  There are no significant differences between the median motions of 1G and 2G stars.}. Clearly the two populations exhibit similar rotation patters.


 Panel d of Figure~\ref{fig:rotazione} shows the average position of the stars in the various circular sectors relative to the cluster center. As expected, 2G stars have smaller radial distances than 1G stars, as a consequence of the fact that the second generation is the most centrally concentrated (Norris \& Freeman 1982; Milone et al.\,2012; Cordero et al.\,2014).  The arrows, which correspond to the average two-dimensional velocity vector calculated in each sector, indicate the motion of 1G and 2G stars relative to the cluster center in 250,000 years. The fact that the vector directions are within a few degrees of the tangential directions confirms that 47\,Tuc exhibits a significant rotation on the plane of the sky (Anderson \& King 2003; Bellini et al.\,2017).
  We note that the perspective effects induced by the large apparent size on the sky of the cluster and its spatial motion (e.g.\,van de Ven et al.\,2006, see their equation 6) would not affect the tangential component of the motion but will result in an apparent expansion of the cluster. However, this effect is small for 47\,Tuc, which has a slow motion along the line of sight. Moreover, we note that such phenomenon would affect in the same way the two populations and would not affect the conclusion of this work. 
 
To investigate the rotation of 1G and 2G stars as a function of the radial distance from the cluster center, we divided the field of view in various circular annulii. Each annulus is defined by using the method of the naive estimator (Silverman 1986) in the region with radial distance from the cluster center between 0.8 and 18.0 arcmin. Specifically, we defined a series of points separated by a distance of  $d=$2.5 arcmin. The bins are defined over a grid of points, which are separated by steps of $d/2$ in distance. 
 For each bin we calculated the tangential velocity of 1G and 2G stars and estimated the corresponding error as $\sigma_{\rm 1G (2G)}/\sqrt{N_{1G (2G)}-1}$, where $\sigma_{\rm 1G (2G)}$ is the tangential-velocity dispersion of 1G (2G) stars and $N_{\rm 1G (2G)}$ the number of 1G (2G) stars  in each annulus.

\begin{centering} 
\begin{figure} 
  \includegraphics[width=8.cm]{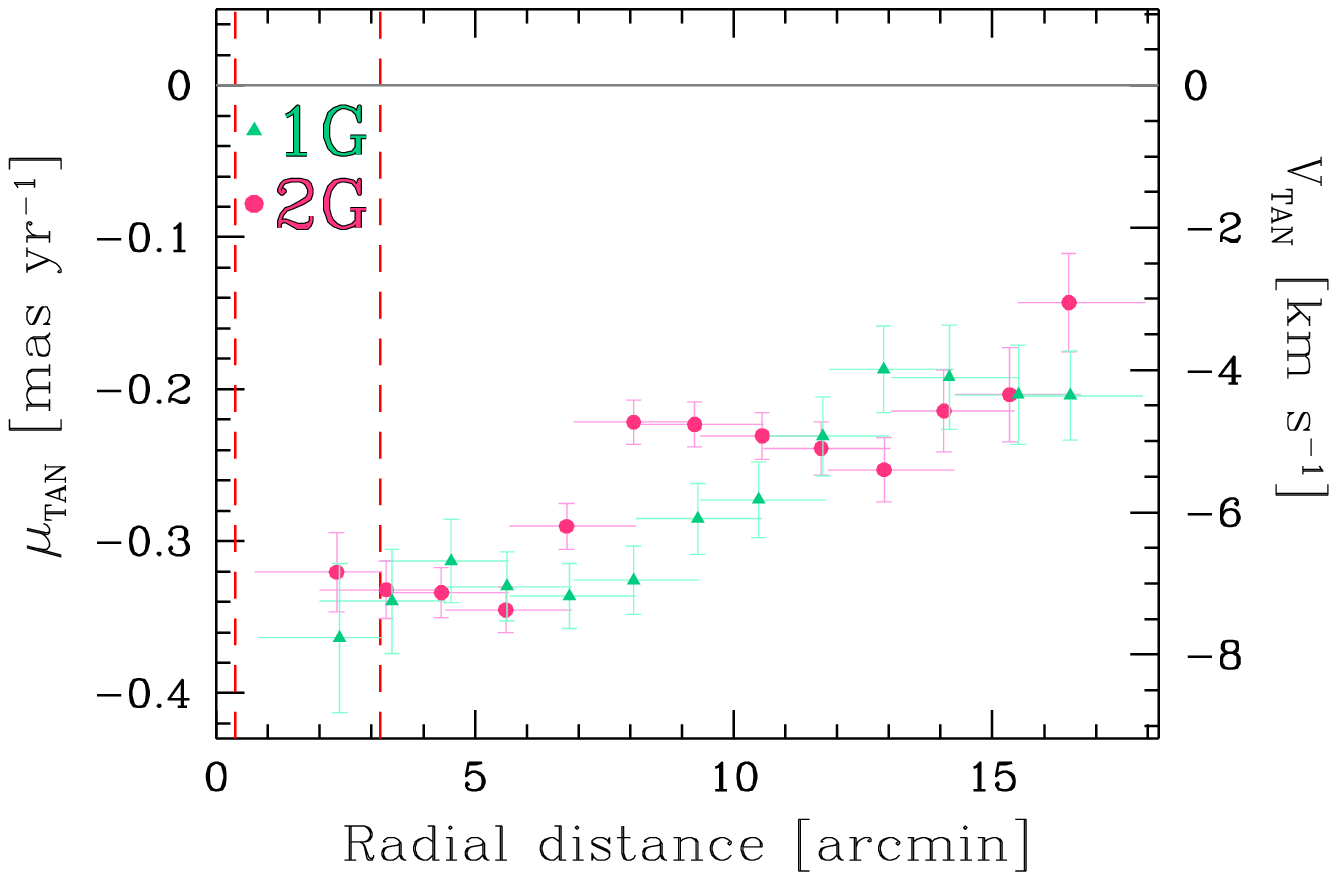}
  \caption{The tangential velocity, which is indicative of the cluster rotation, is plotted as a function of the radial distance from the cluster center. Aqua triangles and magenta dots refer to 1G and 2G stars, respectively. The dashed vertical lines mark the core and the half-mass radius. The scales on the left and right, indicate the tangential velocity in angular and linear units. The latter is calculated by assuming for 47\,Tuc a distance of 4.5 kpc (Harris 1996, updated as in 2010).}
 \label{fig:tan} 
\end{figure} 
\end{centering}

 Our results, illustrated in Figure~\ref{fig:tan}, reveal that 1G and 2G stars exhibit similar rotation along the plane of the sky. 1G stars seem to exhibit slightly higher tangential velocity than 2G stars in the region between $\sim$7
 and $\sim$11 arcmin, although the difference is significant at the $\sim$2-sigma level only.

\subsection{Velocity dispersion of multiple populations}
\begin{centering} 
\begin{figure} 
  \includegraphics[width=8.0cm]{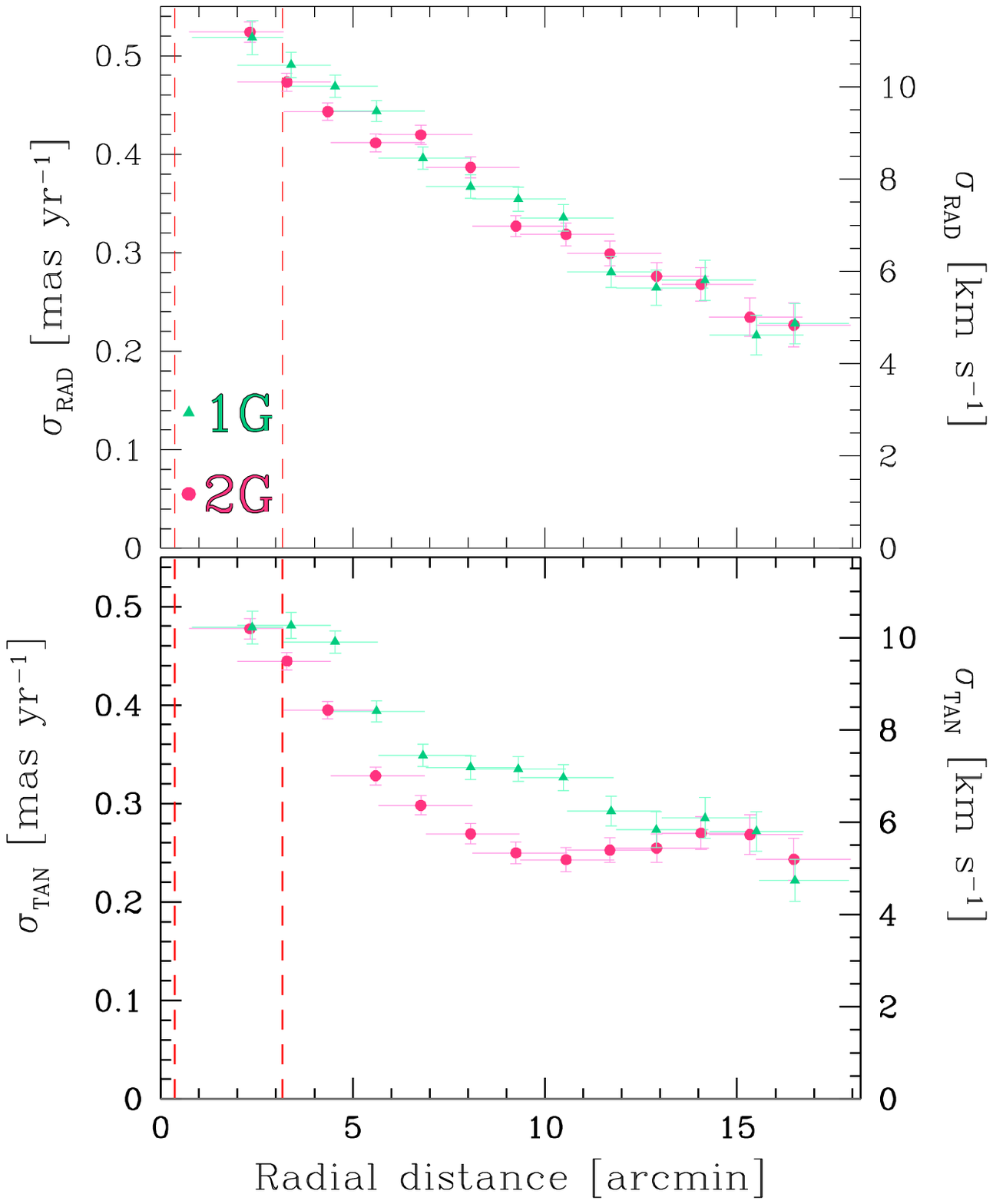}
  \caption{Tangential velocity dispersion (upper panel) and radial velocity dispersion (lower panel) as a function of the radial distance from the cluster center for 1G (aqua triangles) and 2G (magenta circles) stars.}
 \label{fig:disp} 
\end{figure} 
\end{centering}

We calculated the radial ($\sigma_{\rm RAD}$) and the tangential velocity dispersion ($\sigma_{\rm TAN}$) of 1G and 2G stars in each radial bin defined in the previous subsection and plot the corresponding velocity-dispersion profiles in Figure~\ref{fig:disp}.
 To derive the dispersions we adapted to 47\,Tuc, the procedure described in Mackey et al.\,(2013) and Marino et al.\,(2014) that accounts for the contribution of observational errors to the proper-motion dispersion. Briefly, we used a maximum-likelihood technique, assuming that the stellar proper motions are normally distributed around the average value according to their
measurement uncertainties and the intrinsic dispersion. We estimated the intrinsic dispersion by maximizing the logarithm of the joint probability function for the observed proper motions.
 The uncertainties associated to each point are determined by bootstrapping with replacements performed 1,000 times. The error bars indicate one standard deviation (68.27$^{\rm th}$ percentile) of the bootstrapped measurements.

We find similar radial-velocity dispersion profiles for 1G and 2G stars and no significant difference between the radial-velocity dispersion of 1G and 2G stars in the analyzed radial interval.
 In contrast, 1G stars exhibit, on average, smaller tangential-velocity dispersion than the 2G ones. Such difference seems to increase when moving from the half-light radius towards the outermost cluster region and is maximum for radial distance of about 10 arcmin from the cluster center. The tangential-velocity dispersion of 1G stars is consistent with that of 2G stars at radii larger than $\sim$12 arcmin\footnote{Bellini et al.\,(2017) used {\it HST} images to measure the rotation of 47\,Tuc in the plane of the sky and the velocity anisotropy profile from the cluster core out to about 13 arcmin. Although, our paper is focused on the relative motions of 1G and 2G stars and not on the overall cluster dynamics, we verified that the tangential-velocity profiles by Bellini and collaborators are consistent with those of our paper at $\sim$0.5-1.5 $\sigma$ level. The velocity dispersion average profile derived for 1G and 2G stars in this paper are in agreement with those provided by Bellini and collaborators within one $\sigma$. For radial distances larger than $\sim$10 arcmin from the cluster center both the tangential and the radial velocities dispersions derived by Bellini and collaborators are consistent with those derived in our paper at $\sim$1.5-$\sigma$ level with those of our paper, with Bellini et al. providing higher dispersion values. The investigation of such small discrepancy is beyond the purposes of our paper.}.

\begin{centering} 
\begin{figure} 
  \includegraphics[width=8.0cm]{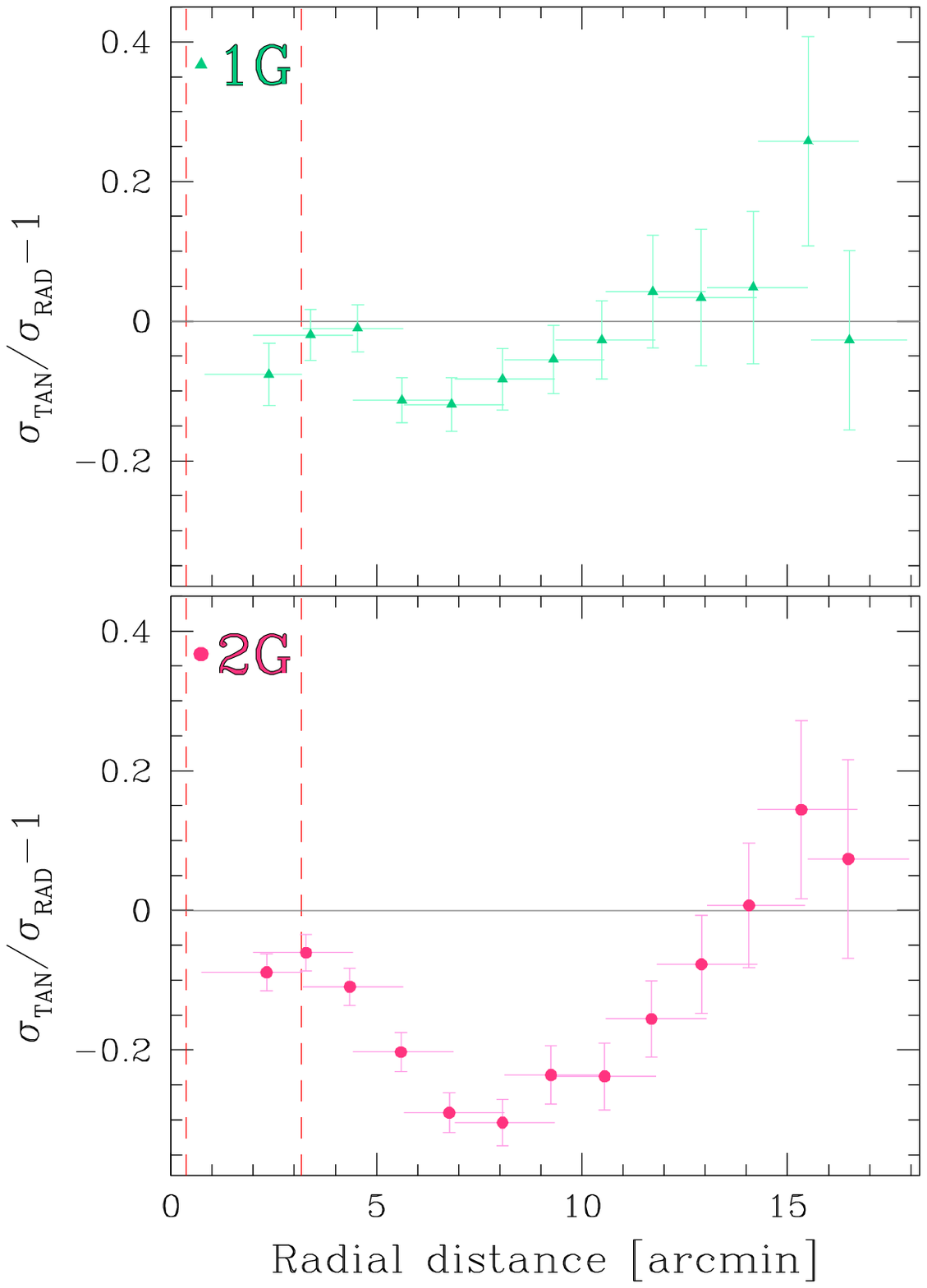}
  \caption{Tangential to radial isotropy for 1G stars (top) and 2G stars (bottom) against the radial distance from the cluster center.}
 \label{fig:anisotropia} 
\end{figure} 
\end{centering}

To further investigate the motion of 1G and 2G stars, we calculated for each population the quantity $\sigma_{\rm TAN}/\sigma_{\rm RAD}-1$, which is indicative of deviation from the isotropy. The radial dependence of $\sigma_{\rm TAN}/\sigma_{\rm RAD}-1$ is illustrated in Figure~\ref{fig:anisotropia} where the horizontal dotted lines correspond to an isotropic stellar system.
2G stars significantly deviate from isotropy in the analyzed region with radial distance from the cluster center smaller than $\sim 12$ arcmin. The value of $\sigma_{\rm TAN}/\sigma_{\rm RAD}-1$ decreases from $\sim -0.1$ to less than $\sim -0.2$ when moving from the half-light radius to about 12 arcmin from the center. This quantity increases in the cluster outskirts, where it is consistent with zero.

The first generation exhibits a mild deviation from isotropy in the region with radial distance from the center between $\sim$5 and 8 arcmin, where the values of $\sigma_{\rm TAN}/\sigma_{\rm RAD}-1$ are smaller than zero and this difference is significant at $\sim$1.5- 2.0-$\sigma$ level. 1G stars are consistent with an isotropic system at radial distances larger than 7 arcmin.
In the outermost bin both 1G and 2G stars have slightly positive $\sigma_{\rm TAN}/\sigma_{\rm RAD}-1$ but the difference from zero has low statistical significance.

\section{Summary and Discussion}\label{sec:summary}

In the last years, several scenarios of formation of multiple populations in GCs have been suggested (see Renzini et al.\,2015 and references therein for a critical review). Some of these scenarios suggest that GCs have experienced at least two bursts of star formation and that 2G stars formed out from material ejected from more-massive 1G stars. Asymptotic-giant branch stars (AGBs), fast-rotating massive stars (FRMSs) and supermassive stars  have been proposed as possible polluters (e.g.\,Cottrell \& Da Costa 1981; Ventura et al.\,2001; Decressin et al.\,2007; Denissenkov \& Hartwick 2014). As an alternative, all the stars in GCs are coeval and multiple populations are the result of exotic phenomena that occurred in the unique environment of the proto GC (e.g.\,De Mink et al.\,2009; Bastian et al.\,2013; Gieles et al.\,2018). 

The various 2G-formation scenarios predict different geometries for the gas from which the 2G form. The signatures left by the evolution of the two nested populations are an exquisite tool to constrain the origin of the younger population.
In this context, 47\,Tuc is an ideal cluster. Indeed, the evidence that its populations are not fully mixed (Norris \& Freeman 1979; Milone et al.\,2012; Cordero et al.\,2014) demonstrates that this cluster has not reached a complete relaxation. As a consequence, it is expected to have retained some of the initial differences between the populations (H{\'e}nault-Brunet et al.\,2015). 

Unfortunately, the dynamical implications of the various scenarios 
  have not yet been fully explored from the dynamical point of view except for few studies that assumed an initial spherical configuration for both populations
  (e.g.\,Decressin et al.\,2010).
  
  So far, the AGB scenario (D'Antona et al.\,2016; D'Ercole et al.\,2008, 2010) is the most developed one in terms of dynamical predictions. In this scenario, the gas produced by slow stellar winds inflows to the centre of the cluster carrying a small amount of angular momentum inherited from the stars of origin, leading to the formation of a 2G gaseous disk (e.g.\,Bekki 2010, 2011).
 Once settled, the disk fragments and forms a 2G stellar disk which relaxes within the 1G component of the cluster. This process alters the initial structure of the system, leading to an elliptical, anisotropic and differentially-rotating GC (see  Mastrobuono-Battisti \& Perets 2013, 2016, H{\'e}nault-Brunet et al.\,2015 and references therein for details). 

  If the relaxation time of the cluster is long enough, the initial phase-space configuration of 2G stars would imprint long-lasting signatures
on the structure of the cluster. As a consequence, the present-day internal dynamics 
  of GCs are directly linked to their past dynamical history.

Recently, Mastrobuono-Battisti \& Perets (2013, 2016) explored the long-term evolution of multiple populations  in the context of the AGB scenario by using {\it N}-body simulations. In the following we qualitatively compare our observational findings with their predictions.

To constrain the internal dynamics of the subpopulations of 47\,Tuc, we first exploited the $V$ vs.\,$C_{\rm U,B,I}$ diagram of stars in this cluster to identify two groups of 1G and 2G stars along the RGB and the HB.
Then we combined photometry and Gaia DR2 proper motions to analyze the rotation and the velocity dispersion along the plane of the sky in the selected 1G and 2G stars.  This approach allowed us to study, for the first time, the internal dynamics of a large sample of more than 3,000 1G and 2G stars over a wide field of view. Specifically, the analyzed region ranges from 0.8 to 18 arcmin from the cluster center, i.e.\,between 0.3 and 5.7 half-light radii.

We discovered that both 1G and 2G stars exhibit a strong rotation on the plane of the sky and that there is no evidence for any difference in the rotation pattern of the two populations.
This finding is apparently in contradiction with the predictions of Mastrobuono-Battisti \& Perets (2016) who find that 2G stars exhibit stronger rotation than the first generation. 
 However, it should be noted that dynamical processes such as relaxation and angular momentum diffusion due to two-body interactions, which act on a short time scale in flattened structures (Haas 2014), could have already acted in the cluster reducing the initial difference (which is of the order of 1-2km/s after one relaxation time), leaving with 1G and 2G stars that rotate at a similar speed. 

We also derived the velocity-dispersion profiles of 1G and 2G stars along the tangential and the radial directions. While there is no significant difference between the radial-velocity dispersion of the two populations, the second generation exhibits, on average, smaller tangential-velocity dispersion than 1G stars. Such difference is more-pronounced in the region with radial distance of about 8-12 arcmin from the cluster center (i.e.\,$\sim$2.5-3.8 times the half-light radius) and strongly decreases when moving towards the innermost or the external regions.
 Second-generation stars strongly deviate from isotropy in the analyzed cluster region with radial distance smaller than $\sim$12 arcmin, in contrast with the first generation, which shows a mild deviation from isotropy at radial distance between $\sim$5 and 7 arcmin. 
 
 Our results are consistent with the conclusion by Richer et al.\,(2013) based on {\it HST} proper motions of stars within a $\sim$3.4$\times$3.4-arcmin region with a distance of 1.9 half-light radii from the cluster center.
 These authors concluded that stars with bluest F606W$-$F814W colors, which likely belong to the 2G, exhibit the largest proper-motion anisotropy which is undetectable for the reddest stars.
 A similar behaviour is observed in NGC\,2808 (Bellini et al.\,2015), $\omega$\,Cen (Bellini et al.\,2017) and, likely, in NGC\,362 (Libralato et al.\,2018), where the result is significant at the $\sim 2 \sigma$-level. These facts would suggest that the high radial anisotropy is a common feature among GCs.

Noticeably, simulations by Mastrobuono-Battisti \& Perets (2016) predict a large difference in the tangential component of the velocity dispersion of the two populations, which is qualitatively similar to what we observe in 47\,Tuc. On the other hand, difference radial-velocity dispersion between 1G and 2G stars is also expected, in contrast with what we observe. 
According to Mastrobuono-Battisti and Perets, the stronger radial anisotropy that characterizes the 2G stars, combined to the presence of rotation, 
 is consistent with the spatial diffusion of a second generation that formed centrally concentrated in a disk configuration.
As an alternative suggested by the referee, the 2G could born in a spherical-centrally-concentrated configuration in a cluster primordially rotating.
 
 It is worth noting that, although our results partially match the predictions by Mastrobuono-Battisti \& Perets (2016), 47\,Tuc is significantly different from the cluster modeled by these authors in terms of mass, relaxation time, and fraction of 2G stars.
 In conclusion, although a more detailed model, tailored to 47\,Tuc will be necessary to properly interpret our results, the internal dynamics of this cluster inferred from Gaia DR2 suggest that the second generation formed in a disk-like, centrally concentrated configuration inside the 1G component. Qualitatively, this work demonstrates how Gaia can contribute to constrain the formation scenarios of the still-eluding multiple populations in GCs.
 
\section*{acknowledgments} 
\small
We thank the anonymous referee for several suggestions that improved the quality of the manuscript.
We are grateful to Peter Stetson who provided the photometric catalog used in the paper. AFM acknowledges support by the Australian Research Council through Discovery Early Career Researcher Award DE160100851.
This work has been supported by the European Research Council through the ERC-StG 2016 project 716082 `GALFOR' (http://progetti.dfa.unipd.it/GALFOR) and by the MIUR through the FARE project R164RM93XW ‘SEMPLICE’. AMB acknowledges support by Sonderforschungsbereich (SFB) 881 ‘The Milky Way System’ of the German Research Foundation (DFG). 
\bibliographystyle{aa}

\end{document}